\documentclass[11pt]{article}

\usepackage{amsfonts}
\usepackage{amssymb}
\usepackage{epsfig}
\usepackage{times}
\newcommand{\be}{\begin{equation}}
\newcommand{\ee}{\end{equation}}
\newcommand{\ba}{\begin{array}}
\newcommand{\ea}{\end{array}}
\newcommand{\bea}{\begin{eqnarray}}
\newcommand{\eea}{\end{eqnarray}}

\def\ket{\,\left\vert  \right\rangle}
\def\bra{\left\langle\right\vert\, }

\def\d{\rm d}

\def\hbar{\not{\hbox{\kern-2.3pt $h$}}}
\def\psl{\not{\hbox{\kern-2.3pt $p$}}}
\def\Psl{\not{\hbox{\kern-2.3pt $P$}}}
\def\ksl{\not{\hbox{\kern-2.3pt $k$}}}
\def\qsl{\not{\hbox{\kern-2.3pt $q$}}}
\def\slad{\not{\hbox{\kern-2.3pt $\partial$}}}
\def\I{\rm i}
\begin{document}
\begin{titlepage}

 \hfill PAR-LPTHE 00/12
\vskip 4.5cm
{\baselineskip 17pt
\begin{center}
{\bf  SUPERHEAVY MAJORANA NEUTRINOS EFFECT IN THE LEPTON-NUMBER VIOLATING  $\;e^- + e^- \to \mu^- + \mu^-$  PROCESS}
\end{center}
}

\vskip .5cm

\centerline{
X. Y. Pham 
     \footnote[1]{E-mail: pham@lpthe.jussieu.fr}
            }
\vskip 2mm
\centerline{{
\em Laboratoire de Physique Th\'eorique et Hautes Energies, Paris}
     \footnote[2]{LPTHE tour 16\,/\,1$^{er}\!$ \'etage,
          Universit\'e P. et M. Curie, BP 126, 4 place Jussieu,
          F-75252 PARIS CEDEX 05 (France).}
}
\centerline{\em Universit\'es Paris 6 et Paris 7;} \centerline{\em Unit\'e associ\'ee au CNRS, UMR 7589.}
\vskip 1.5cm
{\bf Abstract:} In the minimal extension of the standard electroweak theory with ultra massive Majorana neutrinos, the process
$\;e^- + e^- \to \mu^- + \mu^-$ could be observable, in sharp contrast with the reaction $\;e^- + e^- \to $ W$^- +$ W$^-$ which is entirely controlled by neutrinoless double  beta decay $\beta \beta_{0 \nu} $ data. 

Our result provides the   process background  that must be confronted  "new physics" models which postulate doubly charged particles, such as the gauge bilepton $Y^{--}$ in the  $\,SU(3)_{\rm c} \times SU(3)_{\rm L} \times U(1)$ model, the left-right $\,SU(2)_{\rm L} \times SU(2)_{\rm R} \times U(1)_{\rm B-L}\,$ one, and its supersymmetric version with doubly charged Higgs multiplets.  

\smallskip

{\bf PACS} number(s): 14.60.St\quad 11.30.Hv \quad 13.10.+q\quad 12.60.-i
\vfill
\end{titlepage}
%
%

With nonzero neutrino mass recently reported by the Super-Kamiokande collaboration\cite{SK}, it is expected that new physics beyond the  Standard Model (SM) should soon show up in experiment, a typical example would be the observation of rare processes especially those which are absolutely forbidden in the SM. The amply discussed\cite {Fram} electron-electron  option of the future next linear collider (NLC) is an  interesting area for investigating new physics. It provides the prospect for the discovery of L$_{\rm e}$, L$_\mu$,  L$_\tau$ lepton-number violation, especially the  Dirac versus Majorana nature of neutrinos, their masses and mixing. 

 While the first question -- are neutrinos massless or massive?-- is presumably 
settled\cite{SK}, the second question on the neutrinos nature -- are they Dirac or Majorana particles ? -- remains poorly known. The purpose of this note is to point out that the $\;e^- + e^- \to \mu^- + \mu^-$ reaction could open a new window to answer the second question, in competition with  $\beta \beta_{0 \nu} $, the "gold-plated" neutrinoless double beta decay of nuclei $(A, Z) \to (A, Z+2) + e^- +e^-$ frequently discussed in the 
literature.  We  show that the two processses $\beta \beta_{0 \nu} $ and $\;e^- + e^- \to \mu^- + \mu^-$ are complementary, each one separately provides distinctive constraints to the Majorana nature of neutrinos, their  masses and mixing. Therefore their observations  could give two independent informations; both processes when considered together  could further amplify our understanding in the nature and origin of neutrino masses, their  Dirac or  Majorana component.  This $\;e^- + e^- \to \mu^- + \mu^-$ reaction is {\it independent} of  $\beta \beta_{0 \nu} $, in sharp contrast to  the $\;e^- + e^- \to $ W$^- +$ W$^-$ process which is  {\it directly related}\cite{Be} to $\beta \beta_{0 \nu} $. It is important to realize that while observation of $\beta \beta_{0 \nu} $ decay cannot be directly translated to a value for the neutrino mass, it can certainly be used to infer the existence of a non-vanishing Majorana mass regardless of whatever mechanism causes $\beta \beta_{0 \nu} $ to occur\cite{MoPal}. Precisely, if $\beta \beta_{0 \nu} $  is {\it not} seen at a certain level, its absence does {\it not} imply an upper bound on Majorana neutrino mass, but if it {\it is} seen, its presence {\it does} imply a nonzero lower bound on  neutrino mass\cite{Bk}. As we will see, these
 basic facts equally apply to the reaction $\;e^- + e^- \to \mu^- + \mu^-$ that
 we are discussing now. 

In the minimal extension of the $SU(2)_{\rm L} \times U(1)_{\rm Y}$ Standard Theory with  massive Majorana neutrinos, four Feynman diagrams contribute to the   $\;e^- + e^- \to \mu^- + \mu^-$ scattering in the most general renormalizable $R_\xi$ gauge. It is important to note that only massive  Majorana neutrinos,  but not massive Dirac neutrinos, can give rise to the $\;e^- + e^- \to \mu^- + \mu^-$ 
reaction. The  box  with two left-handed W$^-$ gauge bosons exchanged  shown in Fig.1 is one  graph. The three others not shown here are similar to Fig.1 in which the   W$^-$ is replaced in all possible ways by the unphysical Goldstone $\phi^-$ boson, the one absorbed by the W$^-$ to get mass from the Higgs mechanism. Separately each of the four diagrams is $\xi$-dependent, only their sum is gauge-independent. The vertices $\ell N$W and $\ell N \phi$ are given  for instance in\cite{QY} where $\ell$ stands for the electron, muon or down-type quarks and $N$  the heavy neutrino fields or up-type quarks.
The following identities are useful when dealing with Majorana neutral fermions:
\be
 \overline{\ell} \gamma_\mu (1\mp \gamma_5) N = - \overline{N^c}\gamma_\mu (1\pm \gamma_5) \ell^c \;\;\;\;\;\;\;{\rm and}\;\;\;\;\; \overline{\ell} (1\mp \gamma_5) N = + \overline{N^c}(1\mp \gamma_5) \ell^c \;,
\ee 
where $\ell^c$ and $N^c$ are respectively the charge-conjugate of the $\ell$ and $N$ fermionic fields, generically denoted by $\psi$ with $\psi^c = C \overline{\psi}^t$ 
and $ C^{-1} \gamma_\mu \,C = -\gamma_\mu^t$. For Majorana field $N_{\rm maj}$, one has $N_{\rm maj}^c =\eta^* N_{\rm maj}$ where $\eta$ is the phase creation factor of the field $N_{\rm maj}$.

As an illustration, let us explicitly write down  the $\;e^- + e^- \to \mu^- + \mu^-$ amplitude given by the  diagram of Fig.1, 
neglecting the external momenta (see however the remark below) and using the Feynman-'t Hooft $\xi=1$ gauge:
\be
\int {\d^4 k \over (2 \pi)^4} [ \overline{\mu}\gamma^\lambda (1-\gamma_5) {\I\over \ksl -M_i}\gamma^\rho (1+\gamma_5) \mu^c ]\;
 [ \overline{e^c}\gamma_\rho (1+\gamma_5){\I\over \ksl -M_j} \gamma_\lambda (1-\gamma_5) e ] {(-\I)^2 \over (k^2-M_W^2)^2 } \;,
\ee 
the factor $\left({-\I g \over 2 \sqrt{2}}\right)^4 
\left(U_{\mu i} U_{e j}^*\right)^2 \eta_i \eta_{j}^*$ is implicitly included.
Here $U_{\ell i}$ denotes the  Maki--Nakagawa--Sakata (MKS) mixing matrix in the lepton sector, the analog of the Cabibbo-Kobayashi-Maskawa (CKM) of the quark sector. The neutrino 
mass is  $M_i$ and $g^2/8M_{\rm W}^2 = G_{\rm F}/\sqrt{2}$ with $G_{\rm F}\approx 1.166\times 10^{-5}/$(GeV)$^2$ being the Fermi coupling constant. 

The sum of the four diagrams yields the final result for the $e^-(p) +e^-(p') \to \mu^-(P) +\mu^-(P')$ amplitude denoted by ${\cal A}$,
\be
{\cal A} = {\cal K} {G_{\rm F} \over \sqrt{2}} {\alpha_{\rm em} \over 2\pi
 \sin^2\theta_{\rm W}} \; \left[\overline{u}(P)(1+\gamma_5) v(P') + 
P\leftrightarrow P'\right] \otimes\left[\overline{v}(p)(1-\gamma_5) u(p') +
p\leftrightarrow p'\right] \;,
\ee 
where the identity
$$[\overline{u}\gamma^\lambda\gamma^\rho (1+\gamma_5) v] \otimes \left[\overline{v}\gamma_\rho\gamma_\lambda(1-\gamma_5) u \right] = 4 \left[\overline{u} (1+\gamma_5) v\right] \otimes \left[\overline{v}(1-\gamma_5) u \right] $$
has been used.
 
The coefficient ${\cal K}$ which encapsulates all the dynamics due to virtual Majorana neutrinos in the loops is found to be
\be
{\cal K} =  \sum_{i,j}\eta_i \eta_{j}^* \left(U_{\mu i} U_{e j}^*\right)^2 \sqrt{x_ix_j}\left[\left(1+{x_ix_j\over 4}\right) F(x_i,x_j) +{G(x_i,x_j)\over 2} \right]\;\;{\rm with}\;\; x_{i, j}={M_{i,j}^2\over M_{\rm W}^2} \;,
\ee 
and
\be
F(x,y) = F(y,x) = {1\over x-y} \left[ {x\ln x\over (x-1)^2}
 - {y\ln y\over (y-1)^2} +{x-y\over (x-1)(y-1)}\right]\;,
\ee 
\be
G(x,y) = G(y,x) = {1\over x-y} \left[ {x^2\ln x\over (x-1)^2}
 - {y^2\ln y\over (y-1)^2} +{x-y\over (x-1)(y-1)}\right]\;.
\ee 
The typical common factor $\sqrt{x_ix_j}$ in (4) which reflects the Majorana neutrino effect comes from the product $$\left[\gamma^\lambda (1-\gamma_5) {\I\over \ksl -M_i}\gamma^\rho (1+\gamma_5) \right]\left[ \gamma_\rho (1+\gamma_5){\I\over \ksl -M_j} \gamma_\lambda (1-\gamma_5)\right]$$ in (2). The first term $F(x_i, x_j)$ in the bracket of (4)  comes from   W$^-$ W$^-$ exchanged in Fig.1, the second term $x_ix_j F(x_i,x_j)/4$ from $\phi^-\phi^-$ exchange and $G(x_i,x_j)/2$ from W$^-\phi^-$ + $\phi^-$W$^-$.
 When Majorana neutrinos are involved\cite{Bk}, note the amusing fact that electron and muon can be described by the unusual spinor $\,v\,$ rather than the standard spinor $\,u\,$. Also we have
\be
F(x,x) = {(x+1)\ln x - 2(x-1)\over (x-1)^3} \;, F(x,0) = 
{1-x +\ln x \over (x-1)^2}\;, F(1,1) ={1\over 6}\;,F(1,0) ={1\over 2}.
\ee 
\be
G(x,x) = {1-x^2 +2x\ln x \over (1-x)^3} \;, G(x,0) = 
{1-x +x\ln x \over (x-1)^2}\;, G(1,1) ={1\over 3}\;, G(1,0) ={1\over 2}.
\ee 

The corresponding cross section is

\be
{{\rm d}\sigma \over {\rm d}\cos \theta_{\rm cm}} = |{\cal K}|^2
 {\alpha^2_{\rm em}\over 8 \pi^3 \sin^4\theta_{\rm W}} G_{\rm F}^2 s \;\;,\;\sigma = |{\cal K}|^2
 {\alpha^2_{\rm em}\over  4\pi^3 \sin^4\theta_{\rm W}} G_{\rm F}^2 s \;,
\ee 
 where $ s= 4E^2 = (p+p')^2 = (P+P')^2$ is the total energy squared. With identical muons in the final state, the factor ${1\over 2}$ is included in the cross section.

{\bf Remark-}
 In the box diagram calculation given above, only W boson as well as heavy Majorana neutrino masses are kept, while the external momenta are neglected. This approximation turns out to be not unreasonable for two reasons. First, for nonzero $p, p', P, P'$, explicit calculation can be done as it was performed\cite{QY} previously  in a different context, resulting in  a complicated analytic formula with dilogarithm (Spence function) involved, instead of a simple logarithm in (5), (6). This 
approximation is equivalent to  an expansion of ${\cal K}$ as a serie  $\sum_n c_n (s/M_W^2)^n$ for $s < M_W^2$. The leading term $c_0$ is given by (4), the coefficient $c_1$ of the first 
$s/M_W^2$ term is easily computed and  similar to $c_0$ in their analytic expressions. For any fixed $x_i,x_j$, with $\sqrt{s} \approx 100$ GeV, numerically  ${\cal K}$ (without the factor $\eta_i \eta_{j}^* (U_{\mu i} U_{e j}^*)^2$) is damped by a factor of $\approx 1.5$ compared to (4). For $\sqrt{s}= 3 M_W$, we find numerically that $|{\cal K}|$ in (4) decreases by a factor of six, similarly 
to the case found in\cite{QY} for the 
same kinematical $s/M_W^2$ value.

Second, in contrast to the $\;e^- + e^- \to $ W$^- +$ W$^-$ reaction where $s >4M_W^2$ is large, the $\;e^- + e^- \to \mu^- + \mu^-$ is free of this experimental constraint and $\,s\,$ can be small so that the expansion in $s/M_W^2$ makes sense. Note the important fact that in (9), the linear dependence of the cross section on $\,s\,$  is simply due to the kinematical factor $\left[\overline{u}(P)(1+\gamma_5) 
v(P') + P\leftrightarrow P'\right] \otimes\left[\overline{v}(p)(1-\gamma_5) u(p') +
p\leftrightarrow p'\right]$ in (3). This linear $s$ dependence of $\sigma$ tells us that the signal $\;e^- + e^- \to \mu^- + \mu^-$ may be observable at high energies due to kinematics; the dynamics in contained in ${\cal K}$ $\bullet$

For  heavy Majorana neutrinos $x_i \ge 1$, the most important contribution to ${\cal K}$ comes from the 
$x^3 F(x,x)$ term due to  $\phi^-\phi^-$ exchange 
which illustrates the non decoupling 
of heavy fermions in electroweak interactions, unlike QED or QCD. With $x_i \ge 1$ and for not too small mixing, i.e. $\eta_i \eta_{j}^* (U_{\mu i} U_{e j}^*)^2) \approx 10^{-1}, 10^{-2}$,
the coefficient ${\cal K}$ could be of order ${\cal O}(1)$. The cross section $\,\sigma$ which linearly increases with $s$ could get a few femtobarn ($10^{-39}$ cm$^2$) at $\sqrt{s} \approx 100$ GeV. This indicates a readily detectable $L_{e,\mu}$ violating
 event in a collider with integrated luminosity of $10^{33} /$cm$^2$s for a few months. We also note an important fact that  the masses $M_i$ and the mixing $U_{\ell i}$ are in general {\it independent} each other. The assumed behaviour $U_{\ell i} \sim (m_{\ell}/M_i)^k$ with $k >0$, inspired from the empirically observed CKM matrix in the quark sector, is necessarily model-dependent\cite{Kuo}.

As already noted in\cite{Fram}, the important fact is that unlike the  $\;e^- + e^- \to $ W$^- +$ W$^-$ process, the dynamical coefficient ${\cal K}$ in $\;e^- + e^- \to \mu^- + \mu^-$
is not  {\it directly related} to the following relevant quantities in $\beta\beta_{0\nu}$ decay 
which are either
\def\LL{\bra\nu \ket_{\rm L} }
\def\HH{\bra\nu^{-1} \ket_{\rm H} }
\def\qq{\bra q^2 \ket }
\be
\sum _j {\eta_j |U_{e j}|^2 M_j \over q^2}
\ee  
for  light Majorana neutrinos ($M_j^2\ll q^2$) where $q \approx 100$ MeV is the momentum transfer between nuclei, or 
\be
 \sum _j  {\eta_j |U_{e j}|^2\over  M_j} 
\ee 
for 
heavy Majorana neutrinos ($M_j^2\gg q^2)$.

The lastest neutrinoless double beta decay of $^{76}$Ge performed at the Gran Sasso laboratory\cite{Cern} gives
\be
\LL \equiv 
\sum _j \eta_j |U_{e j}|^2 M_j < 0.2 \;{\rm eV}\;,
\ee 
or
\be
\HH \equiv 
\sum _j   {\eta_j |U_{e j}|^2\over  M_j} < 10^{-5}\;{\rm  TeV}^{-1}\;,
\ee

 where the highly nontrivial  nuclear physics matrix element  $\bra q^2\ket$ is taken into account.

Since the neutrino masses $M_i$ and their mixing $U_{\ell i}$ enter differently  in
${\cal K}$ on the one hand and in $\bra\nu \ket_{\rm L} $, $\bra\nu^{-1}\ket_{\rm H}$ on the other hand, the experimental constraints on $\beta\beta_{0\nu}$ cannot be of direct use for
$\;e^- + e^- \to \mu^- + \mu^-$, contrarily to the $\;e^- + e^- \to $W$^{-}$ + W$^{-}$ case\cite{Be}. 
Of course if {\it all} the $M_i, M_j$ are vanishingly small regardless of the mixing elements $U_{e j}$ and $U_{\mu i}$, then both $\beta\beta_{0\nu}$ and
$\;e^- + e^- \to \mu^- + \mu^-$ are desperately unobservable. However this scenario unlikely occurs. If the Majorana neutrinos are light, the small 
$0.2$ eV value in $\bra\nu \ket_{\rm L} $  is presumably due to the cancellation among different terms in (10), each term could have a  larger mass and their mixing could have opposite sign. 

The second scenario with superheavy Majorana neutrinos is equally possible provided that their masses and mixing satisfy the constraint (13) for $\bra\nu^{-1}\ket_{\rm H}$. This happens in grand unified theories and especially in the seesaw mechanism\cite{GMRSY} for which neutrinos with vanishingly small masses as reported  in oscillation experiments are {\it naturally} understood and {\it always accompanied} by superheavy Majorana neutrinos. Here again  different masses $M_j$ ranging from 100 GeV to 10 TeV can accommodate the $\beta\beta_{o\nu}$ data due to the cancellation among different terms in  
$\bra\nu^{-1}\ket_{\rm H}$. Within the constraint (13), our coefficient ${\cal K}$ {\it is not negligibly small} and could be easily of order ${\cal O} (1)$, for instance with $M_1 \approx 100$ GeV and $\eta_1U^2_{e1} \approx 10^{-2}$ while $M_2 \approx 1$TeV and $\eta_2U^2_{e2} \approx -10^{-1}$, just to give a feel for the numbers.

 By comparing (4) with (11), we note the crucial point: the combinations of masses and mixing in ${\cal K}$
 and in $\bra\nu^{-1}\ket_{\rm H}$ are very dissimilar. Moreover the $U_{\mu i}$ is lacking in 
$\bra\nu^{-1}\ket_{\rm H}$ and present in ${\cal K}$, therefore while the constraint of $\beta\beta_{0\nu}$ severely controls $\;e^- + e^- \to \;$W$^{-}$ + W$^{-}$, it has little impact on $\;e^- + e^- \to \mu^- + \mu^-$. We have two independent processes $\beta\beta_{0\nu}$ and $\;e^- + e^- \to \mu^- + \mu^-$, instead of only one with $\beta\beta_{0\nu}$ and $\;e^- + e^- \to \;$W$^{-}$ + W$^{-}$, the latter is, in some sense, redundant.

Our loop background given in (9)  is now  compared with the same process  $\;e^- + e^- \to \mu^- + \mu^-$  governed by the tree diagram bilepton gauge boson $Y^{--}$
 exchange\cite{Fram} of the $\,SU(3)_{\rm c} \times SU(3)_{\rm L} \times U(1)$ model.
The cross section in this model, denoted by $\sigma(Y)$, may be written as
\be
{{\rm d}\sigma (Y) \over {\rm d} \cos \theta_{\rm cm}} = {|\rho|^2 \over 8\pi}\left({M_{\rm W}\over M_{Y}}\right)^4 {1+\cos^2 \theta_{\rm cm} \over (1-s/M_{Y}^2)^2 }
 \; G_{\rm F}^2 s \;\;,\; \sigma (Y) = {|\rho|^2 \over 3\pi}\left({M_{\rm W}\over M_{Y}}\right)^4 { G_{\rm F}^2 s \over (1-s/M_{Y}^2)^2 }
 \; 
\ee 
where $\rho \le 1$  is the electron-muon flavour mixing in the model.
 The factor 
$|\rho|^2 \left({M_{\rm W}\over M_{Y}}\right)^4 \le 10^{-5}$ is  comparable with
$\alpha_{\rm em}^2 |{\cal K}|^2 /\pi^2$ in (9), since ${\cal K}\approx {\cal O}(1)$ due to superheavy Majorana neutrinos involved. Also we note the difference between (9) and 
(14) in their  $\cos \theta_{\rm cm}$ angular distribution and in their $s$ dependence. 

 In conclusion, the  $\;e^- + e^- \to \mu^- + \mu^-$ could be of considerable importance  to reveal the Majorana nature of neutrinos since massive Dirac neutrinos cannot give rise to this reaction. It should be therefore considered on the same footing with the neutrinoless double beta decay. Both processes are equally competitive in their probe of the neutrinos nature, masses and mixing. Contrarily to the low energy nuclei $\beta\beta_{0\nu}$
 decay where superheavy neutrinos imply $1/M_j$ behaviour from (11), the $\;e^- + e^- \to \mu^- + \mu^-$  is more sensitive to the neutrino mass because of its $M^2_j \ln (M_j)$ dependence, as shown by (4) and (7). Paradoxically, unless nontrivial cancellation should occur in ${\cal K}$ between mixing $U_{\ell i}$ and  
masses $M_i$, the  $\;e^- + e^- \to \mu^- + \mu^-$ cross section could be too large for 
$M_i\gg M_{\rm W}$.  In any way, our result  (9) provides  the inevitable background that must be confronted "new physics" models in the search of lepton-number violation.


\begin{em}

\end{em}

\begin{thebibliography}{50}
%
\bibitem{SK}
 The Super-Kamiokande collaboration, Y. Fukuda et al. Phys. Rev. Lett. {\bf 81}, 1562 (1998); ibid {\bf 81}, 4279 (1998); See also some recent reviews, for instance: S.M. Bilenky, C.Giunti and W. Grimus, Prog. Nucl. Part. Phys. {\bf 43}, 1 (1999); P. Ramond, hep-ph/0001006--0001010; 
Jose W.F. Valle, hep-ph/9911224;
K. Zuber, Phys. Rep. {\bf 305}, 295 (1998).    

\bibitem{Fram}
       P. H. Frampton and A. Rasin,  hep-ph/0002135 and references therein.

\bibitem{Be} T. Rizzo, Phys. Lett. {\bf 116B}, 23 (1982);  C.A. Heusch and P. Minkowski, Nucl. Phys. {\bf 416}, 3 (1994);
G. Belanger, F. Boudjema, D. London and H. Nadeau,  Phys. Rev. {\bf D53}, 6292
 (1996) and references therein; 
P. Duka, J. Gluza and M. Zralek, Phys. Rev.
 {\bf D58}, 053009 (1998) and references therein.

\bibitem{MoPal} Rabindra N. Mohapatra and Palash B. Pal, {\bf Massive Neutrinos in Physics ans Astrophysics}, World Scientific 1991.

\bibitem{Bk} B. Kayser with F. Gibrat-Debu and F. Perrier, {\bf The Physics of  Massive Neutrinos}, World Scientific 1989.

\bibitem{QY} Q. Ho-Kim and X.Y. Pham, Phys. Rev. {\bf D61}, 013008 (2000). 

\bibitem{Kuo} 
E.Kh. Akhmedov, G.C. Branco and M.N. Rebelo, hep-ph/9911364;
T.K. Kuo, G.H. Wu and S.W. Mansour, hep-ph/9912366. 

\bibitem{Cern} Cern Courier January/February 2000; H.V. 
Klapdor-Kleingrothaus and H. Paes, hep-ph/0002109 (2000).

\bibitem{GMRSY} M. Gell-Mann, P. Ramond and R. Slansky in {\bf Supergravity}, edited by D. Freedman and P. van Nieuwenhuizen, North Holland (1979);
Y. Yanagida in {\bf Proc. Workshop Unified Theory and Baryon Number in the Universe},  edited by O. Sawada and A. Sugamoto, KEK (1979).







%
\end{thebibliography}
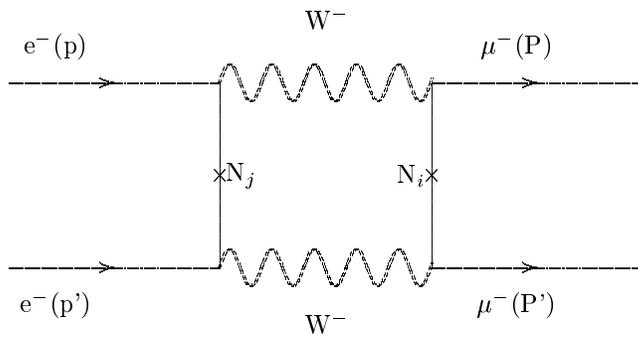
\end{document}